\documentstyle[12pt,epsf]{article}

\def\title{\bgroup\obeylines\everypar={\hskip\parfillskip}\large
	   \bf\vrule height1cm width 0pt\relax}
\def\endtitle{\vskip1sp\egroup}
\def\author#1{\hbox to\textwidth{\hss\vrule height.9cm width0pt\relax
 #1\hss}}
\def\contauthor#1{\hbox to\textwidth{\hss\vrule width0pt\relax #1\hss}}
\def\moreauthors#1{\hbox to\textwidth{\hss\vrule height.8cm
		   width0pt\relax #1\hss}}
\def\instit{\bgroup\small\it\obeylines\everypar{\hskip\parfillskip}}
\def\endinstit{\vskip1sp\egroup}
\def\infap{<\mkern-19mu{\lower1.3ex\hbox{$\sim$}\;}}
\def\supap{>\mkern-19mu{\lower1.3ex\hbox{$\sim$}\;}}
\newcommand{\be}{\begin{equation}}
\newcommand{\ee}{\end{equation}}
\newcommand{\bea}{\begin{eqnarray}}
\newcommand{\eea}{\end{eqnarray}}

\begin{document}
\bibliographystyle{my}

\vglue -4 true cm
\vskip -4 true cm
\def\today{ March 1996}

\begin{center}
{\hfill }{ FTUAM/96-13}
\end{center}

\begin{title}
Do classical configurations produce  Confinement?
\end{title}

\medskip
\begin{center}

A. Gonz\'alez-Arroyo  and A. Montero 

{\it Departamento de F\'{\i}sica Te\'orica C-XI, Universidad Aut\'onoma
 de Madrid,
\\ 28049 Madrid, Spain.}

\end{center}

\begin{center}
\today
\end{center}
\medskip

\begin{abstract}
We show that certain classical SU(2) pure gauge configurations give rise to a
non-zero string tension.  We then investigate cooled configurations generated
by Monte Carlo simulations on the lattice and find similar properties. We infer
evidence in favour of a classical model of Confinement. 
\end{abstract} 	

\newpage

\section{Introduction}
Lattice calculations have given considerable evidence that QCD and  
Yang-Mills theories are confining. In most cases, the reason for having a 
non-zero value of the string tension is hidden under a complicated, numerical 
computation. On the other hand, many authors have attempted  understanding  Confinement in a 
microscopic and qualitative way.  Although it is unclear whether this is possible at all, 
it is certainly desirable, since it provides  more physical insight 
into the origin of the behavior. In the early years of QCD, Polyakov \cite{polyakov}
put forward the idea that classical configurations might play a crucial
role in performing this goal. He showed, that pseudoparticle configurations
(solutions of the euclidean equations of motion) are indeed very important in this 
respect in several theories. It soon became clear, however, that  for QCD, the basic 
pseudoparticle (the instanton) cannot produce Confinement. Instantons seem, nonetheless,
crucial in understanding other properties of the QCD vacuum \cite{thooft,diako,shuryak}. On the 
other hand, Callan, Dashen and Gross \cite{cdg}
proposed that other configurations ---merons--- (singular $Q = \frac{1}{2}$ 
configurations found by De Alfaro, Fubini and Furlan \cite{aff})
could indeed be responsible for this property.  

Quite different approaches to the description of the QCD vacuum and Confinement were taken by other authors.
The Copenhagen group ( for a review see \cite{olesen} ) described the QCD vacuum as a liquid of magnetic 
flux tubes. A discrete version of magnetic flux tubes --- fluxons --- was also  proposed
 as a mechanism for Confinement \cite{Mack}. Fluxons are indeed the relevant disorder configurations 
that cause Confinement for the discrete $Z(N)$ gauge theories. For  other gauge groups, one can define
effective $Z(N)$ degrees of freedom and project out the dynamics for them.  A different 
way to introduce  $Z(N)$ degrees of freedom for a non-abelian gauge theory is through the topology of
boundary conditions. It is remarkable that these twisted boundary conditions (TBC) \cite{twist} were found to be related 
to the configurations of the Copenhagen group.

The most widely accepted description of Confinement is based in the Dual-Superconductor picture 
\cite{thooft1,mandels}. Confinement is seen as  similar to Superconductivity, but with electric and magnetic fields
exchanged. Hence, the condensation of a magnetically charged {\em Higgs} field leads to Confinement. The mechanism is seen to 
work nicely in compact QED \cite{abelian} and in N=2 supersymmetric Yang-Mills \cite{seib}. For N=1 Yang-Mills 
theory, the whole thing is a bit more unclear. One has to explain what is  the magnetic Higgs field (a priori no such field 
is present in the theory) and what is the mechanism 
that causes condensation: One needs the counterpart of the  BCS theory. See Ref \cite{bcs} for interesting ideas in this
direction. 

Recently, a new proposal for understanding Confinement in SU(2) pure gauge
theory  was put forward by our group \cite{lat92,invest}.
The basic idea of the proposal is that a class of quasi-self-dual classical configurations 
produces  Confinement. These configurations are made of self-dual and anti-self-dual regions
 glued together. There is still considerable lack of understanding of how this gluing takes place,
what is the typical size and shape of the regions, etc. The occurrence of self-dual or anti-self-dual pieces patched together
resembles the instanton liquid model of Diakonov, Petrov and Shuryak \cite{diako,shuryak}. However, our main point of 
departure from their  picture is that, in our case, instantons loose their individuallity and dissolve into closely packed multi-instanton
configurations. This introduces considerable difficulty into handling such configurations. Fortunately, we claim to have a
good description of the main bulk of these configurations responsible for Confinement: it can be looked at as a liquid 
of $Q=\frac{1}{2}$ self-dual (or anti-s-d) lumps. This resembles enormously the picture of Ref. \cite{cdg}. However,
our configurations have, to the best of our knowledge, little to do with merons. The configurations we have in mind  are smooth 
and  emerge as deformations of a periodic array of $Q=\frac{1}{2}$ lumps known to exist \cite{marga1} in the presence of
 twisted B.C.'s.  We refer the reader to Ref. \cite{invest} for a more thorough discussion of this scenario. In particular,
we showed there how a scale (mean separation between  lumps, for example) is selected dynamically for the model.
 The approximate value of this scale was extracted from a semiclassical computation to be $0.7 fm$, which was shown to give rise to values for 
the string tension and the topological susceptibility in rough agreement  with measured ones. 

To proceed further, one would like to check the main ingredients of the model: a) there are quasi-self-dual configurations which give rise to a non-zero
string tension;  b) these configurations are arranged into lumps of topological charge $|Q|=\frac{1}{2}$; c) the configurations that dominate the
Yang-Mills vacuum are given by gaussian (and higher) fluctuations around these classical ones.  In Ref. \cite{gauge} we took advantage of the possibility
of checking some of these points by studying  the structure of  lattice configurations. What we actually did, was to generate a collection of fairly 
uncorrelated SU(2) Yang-Mills
configurations at $\beta=2.325$, on an $8^3 \times 64$ lattice with twisted
B.C.'s. Then, in order to get rid of the noise and inspired by the work of Ref. \cite{digiacomo}, we cooled these configurations with the method 
of Ref. \cite{ourcool} and $\epsilon = -0.3$ from $0$ to $50$ cooling steps. Finally, we analyzed the resulting cooled configurations  to 
find local action density maxima which we called {\em peaks}. Our  results were as follows. Indeed, the configurations which one gets after cooling 
are quasi-self-dual and their  total action divided by the number of  peaks is close to $4 \pi^2$, as our model of $Q=\frac{1}{2}$  lumps predicts.
Although we lack a general and neat method to discriminate which  peak is an ordinary instanton ( which should also be present) and 
which is a $Q=\frac{1}{2}$ one, for low-densities we found a clearcut identification method which confirmed that the action per peak result is indeed
due to the presence of a majority of $Q=\frac{1}{2}$ lumps.

To see the relation of these  lumps with Confinement, we computed  the string tension for our configurations. It turns out, 
that the effective string tension, measured from Polyakov loop correlations, decreases with the number of cooling steps. Our main observation
 was that the number of peaks also decreases, due to  instanton-antiinstanton annihilation (which is known to be induced by cooling). Furthermore,
there is a very strong correlation between both decreases: while the string tension varied by more than $40 \%$ from $20$ to $50$ cooling steps,
the quantity $K \equiv \sigma / \sqrt{D}$ only varied by less than $2 \%$ ($\sigma$ is the string tension and $D$ the density of peaks).

Indeed, we claim that the dimensionless variable $K$ is a fundamental quantity for all models of Confinement based on classical configurations.
Since the classical Yang-Mills theory has no scale, if we expand the spatial size by some factor, this should affect densities and square string tension
in the same  way, but $K$ would not change.  Using   the mentioned estimated $ (0.7 fm)^{-4}$ density and the value of $5 fm^{-2}$ 
(taken as normalization) for the string tension, one  gets $K \sim 2.5$, which was consistent with our numerical results.

There are some possible doubts or criticism on our previous work which we will now explain. The first one has to do with the use of TBC's. It could 
be argued, that the presence of $Q=\frac{1}{2}$ objects in our data is an artifact of the use of these B.C.'s, which  stabilize these lumps. Our reply is 
that B.C's  should not affect very much   whenever the size is large enough to contain many lumps, which we claimed was the case.  A different point of criticism, 
could come from our identification of the effective string tension, describing correlations at distances of 5-8 lattice spacings, with the asymptotic
value which signals Confinement. In this respect the work of Teper \cite{teper} cast some doubt on the usefulness of cooling for investigating
the structure of the vacuum. Teper argued that cooling, being a local algorithm, should leave the string tension unaltered. The decrease at fixed
distance of the effective string tension was given no physical significance. Furthermore, he argued then that the true string tension had little
to do with the classical structures revealed by cooling,  and followed from long-wavelength fluctuations of a different nature. However, we can argue back,
that the recovery of the uncooled string tension at much larger distances is not in contradiction with a classical picture, since after all, 
the basic effect  induced by  cooling (the 
annihilation of pairs) is  also local in nature and should produce the same effect.

In this work, we are trying to obtain a stronger and more direct evidence supporting our argumentation.  
\section{Data}
Are there some self-dual configurations which give rise to a non-zero string tension?
To answer this question, we have produced   such configurations and measured Wilson loops on them. In view of 
  our theoretical prejudices, we started with a periodic  array of  $Q=\frac{1}{2}$ instantons. This was 
obtained by taking the  $4^4$ lattice (with twisted boundary conditions) single lump configuration and 
gluing it  to itself  to produce a $16^3 \times 8$ (purely periodic B.C.) configuration. Due to the 
periodic arrangement of the  resulting 128 lumps, this configuration is not suitable by itself. 
To randomize the position of the lumps, we applied a series of Monte Carlo steps to this configuration
using  Wilson action and several values of $\beta$ in the range $[ 3,4] $ in steps of 0.2.
 The number of Monte Carlo steps was taken such that 
the final spatial distribution of peak locations was uniform within errors. For  $\beta=3.0-3.2$, 50 steps was found sufficient, 
while the number had to increase for larger $\beta$.  Finally, to eliminate noise, we  applied up to 50  cooling steps (in steps of 10) with the
algorithm  of Ref. \cite{ourcool} and $\epsilon = -0.3$. For each value of $\beta$, 10 different configurations were generated
in the same way. 
The results are basically consistent in all cases, so  we will illustrate them, by showing   
the numbers for  $\beta=3.2$ and $50$ Monte Carlo steps and $\beta=3.8$ and $75$ Monte Carlo steps. After the first 10-20   cooling steps, 
things change very little, so we will quote   
only the results for 10 and 50 cooling steps. These are given in Table I.

The main characteristics of the data are the following. The configurations are close to self-dual, as can
be seen by comparing the mean action divided by $8 \pi^2$ with the mean topological charge. The Monte Carlo 
steps have, not only randomized the position of the peaks, but also decreased their number down  to the  
20 to 40 range.  However, in agreement with the initial configuration they all have positive topological charge
and an action per peak which is close to $4 \pi^2$. This is also similar to the figure obtained for our TBC data \cite{gauge}.
Next, comes the crucial point: the string tension.

 We measured the average value of $R \times T$ Wilson loops, for $R,T$ going from $1$ to $7$ in all the $16 \times 16$ planes (3 planes),
 over  all the positions of the loops and the 10 independent configurations. 
 From these averages  we extracted the Creutz ratios  $\chi(R,T)$. These numbers, show  
a dependence on the loop size $R\times T$ which  flattens up for larger values. Since exploring larger
loops was limited by our lattice sizes (and this by our computer memory) we tried to understand the main origin of the dependence. 
Creutz ratios are by definition insensitive to perimeter and constant terms in the exponent of Wilson 
loop averages. However,  scale-invariant perturbative-like terms do not drop out 
and  produce a  $\gamma ( 1/(R(R-1)) + 1/(T(T-1)))$ dependence. We henceforth
fitted our  Creutz ratios data to the form:
\begin{equation}
\chi(R,T) =  \sigma  -  \gamma ( \frac{1}{R(R-1)} + \frac{1}{T(T-1)}) 
\end{equation}
The errors on $\chi(R,T)$ were obtained from the dispersion over the 10 configurations. The fits are excellent for all number of cooling steps.
Fitting from $R,T = 4 $ onwards gives $\chi^2$ values which are less than one for 16 degrees of freedom 
and 2 parameters. Fitting from $R,T = 3  $  or $5$  give also good   $\chi^2$ values and consistent results. 
To show the quality of our results, we display  in Fig. 1 those  for $\beta=3.2$
and 50 MC steps. What we actually plot is the combination
\begin{equation}
\sigma(R,T) \equiv \chi(R,T)  + \gamma ( \frac{1}{R(R-1)} + \frac{1}{T(T-1)})   
\end{equation}
as a function of the area $R \times T$. The constancy of the quantity is quite impressive, showing that Creutz ratios
can be very well described by an area law and a perturbative-like piece. From the plot it is also apparent that data of different sizes are 
strongly correlated (which explains the tiny $\chi^2$ values).   
Hence, the errors given by the fit are unrealistic. A safe estimate for the error of $\sigma$ is  $10 \%$. The result is good for all number of cooling steps.
It is also clear from Fig. 1 and Table I,  that  the value of the string tension $\sigma$ decreases with the number of cooling steps, although 
much less strongly than for our TBC data. This was to be expected, since in this case the basic positive-negative
charge annihilation mechanism is almost absent. The resulting K values are not far from those obtained for the  TBC's configurations.

Having established that certain configurations do produce a non-zero string tension, we proceed to the analysis of another crucial point 
for our argumentation: We have argued that the main conclusions of our previous paper \cite{gauge} were not essentially  biased by the 
use of twisted B.C.'s. For that purpose, we have Monte Carlo generated a total of $50$ configurations on a $16^3 \times 8$
lattice with periodic boundary conditions, using Wilson action and $\beta = 2.325$.  Out of them, 30
configurations were generated sequentially, separated by 500 heat bath
sweeps, after some  initial  5000  sweeps. Each one of the remaining 20, are obtained independently, by applying 500 Monte
Carlo sweeps to an initial configuration, which was taken as fully ordered
for 10 configurations and  random for the remaining 10.
We have analyzed independently all three sets of configurations and found
perfect compatibility, so we will henceforth consider the 50 configurations
together. Finally, we applied up to 100 cooling steps (with $\epsilon = -0.3$)
to them, and  performed the same analysis to the cooled configurations as
employed before.  The results are summarized in Table II and Fig. 2.

In this case, both positive and negative topological charge peaks are present
and, hence, the configurations are neither self-dual nor
anti-self-dual. However, notice from Table II, that  the integral of the absolute value of the topological charge density,
labeled $\tilde{S}$, takes values close to  those of  the total action divided
by $8 \pi^2$. This we interpret as a
sign that the configuration is locally close to self-dual or anti-self-dual with regions
connecting both areas. This is confirmed by the self (or anti-self)-duality
of most peaks. Again, we find that the mean action per peak is close to the
$4 \pi^2$ prediction of our model, and to the twisted B.C. results. 

If we now look at the analysis of Wilson loops averages and Creutz ratios,
we find   similar results as before. Fits to Formula 1 are excellent
as seen in Fig. 2, giving  $\chi^2$ values per degree of freedom  smaller than one.  Now, the string tension 
decreases strongly with cooling, dropping by a factor 3.5 from $10$ to $100$ cooling steps.
However,  this variation is correlated with the drop  in the number of peaks, giving  a value of $K$
 which  changes  only by
$20 \%$. We also stress,  that the actual value, close to $2.5$, is in
agreement with our theoretical estimate, the result for TBC's \cite{gauge}
and our self-dual configurations (these need not give exactly the same value). 
We hence see that, despite the change in lattice size, boundary conditions, and the use of different observables to 
extract the string tension, the  main features of our TBC results are nicely verified.
Another interesting point is that, if we convert the string tension to physical units as in Ref. \cite{gauge}, we get for
10 cooling steps a value of $5.21 fm^{-2}$, close to the uncooled infinite volume string tension, taken as normalization to $5 fm^{-2}$. 
 
\section{Conclusions}
 In summary, our results make it clear that the non-thermal self-dual configurations
obtained from the array of $|Q|=\frac{1}{2}$ lumps do produce a non-zero value for the 
string tension. 
On the other hand, thermalized  configurations generated at $\beta=2.325$ and periodic 
boundary conditions show  a similar behavior with values of the quantity $K$ ( string tension 
divided by square root density) which are close to the previous ones, those found for the 
TBC data and the expectations of our model of the Yang-Mills vacuum. 
We comment that our  results are quite economical in what regards computer resources. 
The self-dual data required 24 hours for each value of $\beta$ (10 configurations) in an
HP-9000 730 workstation. The thermalized configurations required 250 hours ( for 50 configurations)
in total.

We also want to mention that, in all cases, the mean action per peak is found to be close to the 
$ 4 \pi^2$ value predicted by the $|Q|=\frac{1}{2}$ liquid model of the vacuum Ref. \cite{invest}.
In this case, however, we lack a method of identifying each peak, but it is quite unnatural to think that,
given the general agreement in all other aspects,  
a similar value to the one found for the twisted boundary conditions results will have a different origin.

One of us (A.G-A) wants to thank D. Diakonov, V. Petrov,  P. van Baal,  J. Verbaarschot  and the rest of participants 
in  the 1995 Trento
workshop on Non-perturbative approaches to QCD 
for useful conversations and the possibility to discuss all ideas at length. 
This work was supported by the CICYT grants AEN93-0693 and the EC network
CHRX-CT93-0132.

\begin{table}
\begin{center}
\caption{ Results for the non-thermal configurations obtained after 50 MC sweeps at  
 $\beta=3.2$ (Conf. $I$) and 75 sweeps at
 $\beta=3.8$ (Conf. $II$), and  $N_{cool}$ cooling steps.
 We give the mean values of the action $S$ in $8\pi^2$ units,
the  topological charge $Q$ and   number of peaks $N_p$. The results of our fit to Creutz ratios for
  $R,T \geq 4$ ($\gamma$ and the string tension  $\sigma$)
are given. The quantity $K \equiv \sigma / (peak \ density)^{\frac{1}{2}}$ is also shown.}
\begin{tabular}{||c||c|c|c|c|c|c|c||}
\hline 
Conf. &  $N_{cool}$ &  $<\frac{S}{8\pi^2}>$  &   $<Q>$  &  $<N_{p}>$ & $\sigma$ & $\gamma$ & K \\ \hline \hline 
$I$ &  10    &  18.8(1.0)   &  15.8(1.2)   & 31.8(2.4) &  0.067 & 0.140 &  2.16 \\ \hline
$I$ &  50    &  15.3(1.0)   &  15.1(1.0)   & 24.3(3.3) &  0.052 & 0.094 &  1.91 \\ \hline 
$II$&  10    &  16.2(0.8)   &  14.5(0.8)   & 24.2(1.8) &  0.056 & 0.105 &  2.05  \\ \hline
$II$&  50    &  14.2(0.8)   &  14.1(0.8)   & 19.9(1.8) &  0.046 & 0.074 &  1.87  \\ \hline  
\end{tabular}
\end{center}
\end{table}

\begin{table}
\begin{center}
\caption{The same as Table I, but for the thermalized configurations at $\beta=2.325$. In this case we give 
$\tilde{S}$ ( the integral of the absolute value of the topological charge density) instead of $Q$.}
\begin{tabular}{||c||c|c|c|c|c|c||}
\hline
$N_{cool} $ &  $<\frac{S}{8\pi ^2}> $  &  $<\tilde S>$  &  $<N_{p}>$ & $\sigma$ & $\gamma$ & K \\ \hline \hline  
  10    &  35.1(0.6)   &  22.7(0.5)   & 77.8(1.1) &  0.124 & 0.272 &  2.54 \\ \hline
 20    &  19.9(0.5)   &  14.2(0.5)   & 38.2(0.7) &  0.093& 0.250 &  2.70 \\ \hline
  40    &  11.8(0.5)   &   9.2(0.5)   & 19.5(0.7) &  0.063  & 0.185 &  2.58 \\ \hline
 100   &   6.1(0.4)   &   5.4(0.4)   & 8.8(0.4)  &  0.035 & 0.107 &  2.11  \\ \hline 
\end{tabular}
\end{center}
\end{table}

\begin{figure}
 \begin{center}
  \leavevmode
   \epsfverbosetrue
   \epsfxsize=250pt
   \epsffile{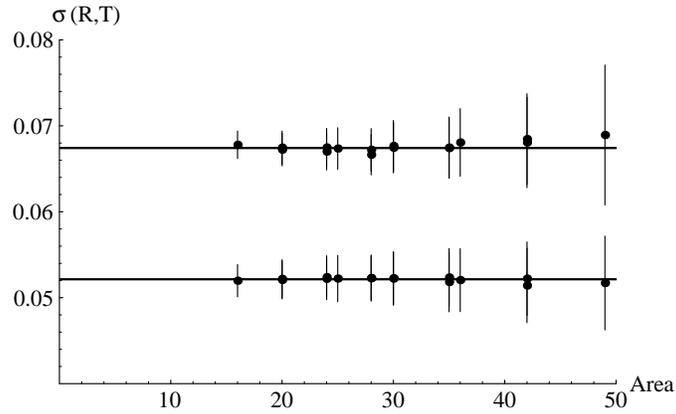}
   \caption{The lattice String Tension approximant $\sigma (R,T)$ (Formula 2) is plotted as a function of the Area $R \times T$ for
 the non-thermal configurations ($\beta=3.2$ and 50 MC steps) after 
         10  (upper points) and 50 (lower points) cooling steps. The horizontal lines are the best 
fit values given in Table I.}
 \end{center}
\end{figure}

\begin{figure}
 \begin{center}
  \leavevmode
   \epsfverbosetrue
   \epsfxsize=250pt
   \epsffile{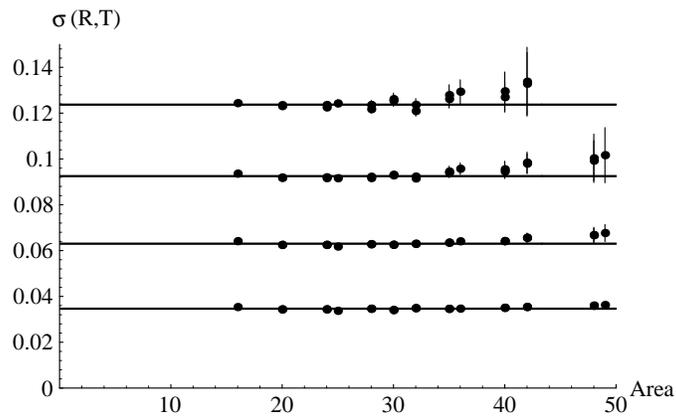}
   \caption{The same as Fig. 1 but for the thermalized data at  $\beta=2.325$. From top to bottom the data for 
    10, 20, 40 and 100  cooling steps is displayed, together with the best fit horizontal lines.}
 \end{center}
\end{figure}

\end{document}